\documentclass[preprint]{article}
\usepackage{authblk}
\usepackage{graphicx}% Include Figure files
\usepackage{dcolumn}% Align table columns on decimal point
\usepackage{bm}% bold math
\usepackage{amsmath}
\usepackage{amssymb}
\usepackage{multirow}
\usepackage{caption}
\usepackage{subcaption}
\usepackage{epstopdf}
\usepackage{hyperref}
\usepackage{cite}
\begin{document}
\title{\bf CPL-parametrized cosmic expansion in Galileon gravity: Constraints from recent data}
\author[]{Amit Samaddar\thanks{samaddaramit4@gmail.com}}
\author[]{S. Surendra Singh\thanks{ssuren.mu@gmail.com}}
\affil[]{Department of Mathematics, National Institute of Technology Manipur, Imphal-795004,India.}

\maketitle

\begin{abstract}
We explore the cosmic expansion history within the framework of Galileon gravity by employing  a redshift-based expression for the Hubble rate, $H(z)$, derived from the CPL parametrization $\omega_{DE}(z)=\omega_{0}+\omega_{a}\frac{z}{1+z}$. This parametrization allows for a time-dependent expansion history consistent with the non-linear Galileon field equations. To constrain the model parameters, we perform a MCMC analysis using $46$ Hubble parameter measurements, DESI DR2 BAO data and $1701$ Pantheon+ datasets. The best fit values obtained  are  $H_0 = 67.7043^{+1.4354}_{-1.4102}$ km/s/Mpc, $\Omega_{m0} = 0.2668^{+0.0212}_{-0.0217}$, $\omega_0 = -0.8827^{+0.1076}_{-0.0967}$ and $\omega_a = 0.0011^{+0.0660}_{-0.0622}$. Model comparison using information criteria yields $\Delta AIC=1.46$ and $\Delta BIC = 11.5$ indicating that the Galileon model is a strong contender to the $\Lambda$CDM model. The deceleration parameter shows a transition at $z_{tr} = 0.7873$, with $q_0 = -0.598$. Energy density and pressure remain physically viable with $\rho_{de}(z)>0$ and $p_{de}(z)<0$ with the present day equation of state $\omega(z)$ value of $-0.2915$, which suggests mild dynamical dark energy. NEC and DECare satisfied, while SEC is violated. The model yields $r_{0}=0.657$, $s_{0}=0.1173$ and  Om diagnostic shows a peak value of $-0.45$ at $z = 0.377$, converging to $-1$ at late times.These results demonstrate that Galileon gravity remains a viable and flexible alternative to $\Lambda$CDM in describing late-time cosmic acceleration.
\end{abstract} 


\section{Introduction}\label{sec1}
\hspace{0.5cm} In recent decades, numerous observational findings have convincingly demonstrated that the Universe is experiencing an accelerated expansion. Initially revealed through measurements of Type Ia supernovae \cite{Riess98}, this surprising discovery has been consistently reinforced by various independent observations, including the cosmic microwave background (CMB) \cite{Planck16}, baryon acoustic oscillations (BAO) \cite{Cole05} and large-scale structure data \cite{Tde15}. Understanding the origin of this accelerated expansion in the late Universe remains a major open problem in modern cosmological research. The prevailing cosmological paradigm, known as the $\Lambda$CDM model, explains the Universe’s late-time acceleration to the presence of a cosmological constant ($\Lambda$) incorporated into Einstein’s theory of general relativity \cite{Swe89,Pad03,Sahni06}. Although this model successfully accounts for numerous observational datasets, it is not without theoretical shortcomings—most notably, the fine-tuning and coincidence problems. These challenges have encouraged the exploration of other theoretical avenues, particularly modified gravity models, which seek to account for the Universe’s accelerated expansion at late times without invoking a cosmological constant.

An important class of modified gravity theories employs scalar fields with extended derivative interactions, carefully formulated so that the equations of motion involve only up to second derivatives. This structure helps to avoid Ostrogradsky instabilities that usually arise in higher-derivative systems. Among these, Galileon gravity stands out as a prominent framework, initially formulated in flat spacetime and later generalized to curved spacetime geometries \cite{Horn74}. The Galileon theory was first formulated by Nicolis, Rattazzi, and Trincherini [3] in flat spacetime as a special class of scalar field models that respect the Galilean shift symmetry: $\phi \rightarrow \phi + c$, $\partial_{\mu} \phi \rightarrow \partial_{\mu} \phi + b_{\mu}$, where $c$ and $b_{\mu}$ are constants \cite{Nicol09}. This symmetry restricts the allowed Lagrangians to a finite number of terms that lead to second-order equations of motion, despite involving higher derivatives in the action. These unique interactions are constructed from scalar combinations of the field and its derivatives, including $\partial_{\mu} \phi$, $\partial_{\mu} \partial_{\nu} \phi$, and higher-order contractions, ensuring theoretical stability and predictive power. The original Galileon framework was later generalized to curved spacetime by \cite{CD09,CD11}, who developed a covariant version of the theory that ensures the equations of motion do not involve derivatives beyond second order. This extension involves introducing specific non-minimal couplings with curvature tensors, allowing the model to remain consistent with general relativity on cosmological scales. Furthermore, it was demonstrated that Galileon terms can emerge as the low-energy effective description in the decoupling limit of higher-dimensional gravity models, such as the Dvali–Gabadadze–Porrati (DGP) scenario \cite{Dvali00}.

One of the hallmark features of Galileon theories is the Vainshtein mechanism \cite{Va72}, a non-linear screening effect that suppresses the scalar field's influence in high-density or small-scale environments. This allows Galileon models to remain consistent with Solar System tests of gravity while exhibiting significant deviations at cosmological scales. Moreover, the Galileon couplings are protected against quantum corrections due to their non-renormalization properties \cite{Hint10}. Cosmologically, Galileon gravity has been explored across a wide range of phenomena. It has been used to model late-time acceleration without a cosmological constant \cite{Ts10}, as well as inflationary scenarios with distinct signatures \cite{Koba11,Ali10,Qiu10}. The theory has also been studied in the context of non-Gaussianities \cite{Burr10,Lang16}, bounce cosmology \cite{Gann10} and structure formation and growth history \cite{Mota10,Tret12}. Galileon perturbations, stability conditions, and spherical solutions have all been thoroughly analyzed \cite{Ger12,Zabat11}, along with various generalizations such as the extended and generalized Galileons, including Horndeski and beyond-Horndeski models \cite{Gley15}. Given its strong theoretical foundation and observational flexibility, Galileon cosmology offers a compelling framework to explore cosmic acceleration beyond the standard $\Lambda$CDM model. By introducing derivative self-interactions of a scalar field while maintaining second-order equations of motion, Galileon models evade Ostrogradsky instabilities and naturally drive late-time acceleration without invoking a cosmological constant. Numerous studies have placed observational constraints on Galileon parameters using a range of cosmological datasets, including Type Ia supernovae (SNe Ia), the CMB, BAO and large-scale structure surveys \cite{Ness16,Gong15,Raveri16,EDV21}. These efforts have facilitated an in-depth analysis contrasting Galileon cosmology with the standard $\Lambda$CDM model, highlighting potential differences in cosmic evolution and structure formation dynamics. In addition to observational investigations, the dynamical behavior of Galileon models has been studied through autonomous dynamical system techniques, providing a qualitative understanding of their cosmological evolution. Such analyses have identified critical points corresponding to radiation, matter, and accelerated epochs, offering insight into the stability and attractor behavior of the solutions \cite{Leon13,Mishra24}. These complementary approaches collectively highlight the robustness of Galileon gravity in capturing both the background expansion and structure formation.

Motivated by the aforementioned studies, we focus on a CPL-parametrized Galileon cosmology, employing the Chevallier–Polarski–Linder (CPL) equation of state for dark energy is expressed as: $\omega(z)=\omega_0+\omega_a \frac{z}{1+z}$, as a phenomenological tool to model the redshift evolution of dark energy. This widely-used parametrization captures possible time evolution in the dark energy EoS without invoking specific scalar field potentials or higher-order corrections, making it ideal for testing the dynamical behavior of modified gravity theories like Galileon models \cite{Amits24,AS25}. In this context, Hubble parameter parametrization refers to modeling $H(z)$, the Hubble expansion rate, in terms of a chosen form of the dark energy EoS. This approach enables a model-independent characterization of expansion history, thereby facilitating direct confrontation with observational datasets like Supernovae Type Ia (SNe Ia), BAO and CMB. In Galileon cosmology, where the scalar field introduces non-trivial modifications to the background dynamics, the use of parametrized Hubble functions helps to isolate the effects of the modified kinetic couplings and derivative self-interactions. Parametrizing the Hubble function via the CPL form is particularly beneficial in Galileon models, as it allows the theory to accommodate both late-time acceleration and tracking behavior without committing to a fixed scalar field evolution. Moreover, it offers a practical framework for performing statistical fits to current observational datasets and comparing the viability of Galileon gravity against $\Lambda$CDM. Our analysis, therefore, aims to constrain the CPL parameters within the Galileon framework and examine deviations from standard cosmology that may be encoded in the Hubble expansion history.

This paper is structured in the following manner. Section \ref{sec2} outlines the formulation of Galileon cosmology, where we derive the corresponding Friedmann equations and explore the theoretical background that supports the model. Section \ref{sec3} focuses on utilizing the CPL parametrization to construct a Hubble parameter formula consistent with the Galileon cosmological scenario. Section \ref{sec4} is dedicated to evaluating the cosmological parameters by confronting the model with observational datasets like Type Ia SNe, BAO features and $H(z)$ datasets. In section \ref{sec5}, we analyze the model using statefinder diagnostics to distinguish it from standard $\Lambda$CDM cosmology and other dark energy models. Section \ref{sec6} is devoted to the Om diagnostic, providing further insight into the deviation behavior of the proposed model from $\Lambda$CDM. To wrap up, section \ref{sec7} outlines the main conclusions drawn from this study and offers closing insights.
\section{Generalized Galileon framework}\label{sec2}
\hspace{0.5cm} This section offers a concise introduction to cosmology within the framework of the fully generalized Galileon theory. We outline the key background equations governing the evolution of the Universe and highlight the criteria required to maintain the theory’s stability \cite{Deff,Felice12}. In order to avoid the emergence of Ostrogradsky instabilities \cite{Ostr50}, which stem from higher-order derivative terms, it is crucial to formulate equations of motion with derivatives not higher than second order. The most inclusive scalar-tensor theories in four-dimensional spacetime that meet this requirement are encapsulated by a Lagrangian constructed as a sum of five fundamental terms:
\begin{equation}\label{1}
\mathcal{L}=\sum_{i=2}^{5}\mathcal{L}_i.
\end{equation}
The generalized Galileon Lagrangian is composed of four core contributions, denoted as $\mathcal{L}_i$ for $i = 2, 3, 4, 5$, each representing a different type of coupling or kinetic structure:
\begin{equation}\label{2}
\mathcal{L}_2= G_2(\phi, X),
\end{equation}
\begin{equation}\label{3}
\mathcal{L}_3= G_3(\phi, X) \Box \phi,
\end{equation}
\begin{equation}\label{4}
\mathcal{L}_4= G_4(\phi, X)(-T + B) + G_{4,X}(\phi, X)\left[ (\Box \phi)^2 - \phi_{;\alpha\beta} \phi^{;\alpha\beta} \right], 
\end{equation}
\begin{equation}\label{5}
\mathcal{L}_5= G_5(\phi, X) G^{\alpha\beta} \phi_{;\alpha\beta} - \frac{1}{6} G_{5,X}(\phi, X) \left[ (\Box \phi)^3 + 2 \phi_{;\alpha}^{\;\;\;\gamma} \phi_{;\gamma}^{\;\;\;\delta} \phi_{;\delta}^{\;\;\;\alpha} - 3 \phi_{;\alpha\beta} \phi^{;\alpha\beta} \Box \phi \right].
\end{equation}
In these expressions, $\phi$ represents the scalar field and the kinetic term is described as $X = -\frac{1}{2} \partial^\alpha \phi \, \partial_\alpha \phi$. The functions $G_i(\phi, X)$ encode the dynamics and self-interactions of the scalar field and their partial derivatives in $\phi$ and $X$ are denoted as $G_{i,\phi}$ and $G_{i,X}$, respectively. These terms collectively generalize the Horndeski class within the teleparallel framework, while preserving second-order field equations to prevent Ostrogradsky instabilities \cite{GWH74}.

Beyond the scalar-tensor contributions discussed above, a comprehensive cosmological model must incorporate the cosmic matter sectors. This is typically achieved by including a matter Lagrangian, $\mathcal{L}_{m}$, which characterizes a uniform fluid component governed by the quantities $\rho_{m}$ (energy-density) and $p_{m}$ (pressure). Accordingly, the full action takes the form:
\begin{equation}\label{6}
S=\int d^{4}x\,\sqrt{-g}\,(\mathcal{L}+\mathcal{L}_m),
\end{equation}
where $g$ denotes the determinant of the spacetime metric $g_{\alpha\beta}$.

It is worth emphasizing an important aspect of this formulation: in the present study, we deliberately omit any direct influence or coupling of the scalar field on the matter distribution. This choice aligns with a common approach adopted in several cosmological investigations \cite{Silva09,Felice10,Shiari12,AAS12,Koba10}. Nevertheless, we acknowledge that the original Galileon framework was conceived with a non-minimal coupling to matter \cite{Nicol09,Esp09,Defff10}, a feature that facilitates mechanisms like the Vainshtein screening. In this study, we consider the Galileon field as an independent dynamical entity that plays the role of dark energy, without interpreting it as an extension or modification of the gravitational sector. We recognize that introducing a coupling between the Galileon and matter could substantially alter the resulting cosmological dynamics and lead to richer phenomenology.

To proceed with the cosmological analysis, we consider a spatially flat Friedmann–Robertson–Walker (FRW) spacetime characterized by the line element:
\begin{equation}\label{7}
ds^{2}=-N^{2}(t) dt^{2}+a^{2}(t)\,(dx^{2}+dy^{2}+dz^{2}),
\end{equation}
where $t$ denotes the cosmic time, $N(t)$ is the lapse function and $a(t)$ represents the scale factor of the Universe. By performing variation of the action (\ref{6}) with respect to the lapse function $N(t)$ and the scale factor $a(t)$, and subsequently adopting the gauge choice $N(t)=1$, we derive the corresponding Friedmann equations that govern the background cosmological evolution.
\begin{align}\label{8}
&2X G_{2,X} - G_2 + 6 X \dot{\phi} H G_{3,X} - 2X G_{3,\phi} - 6 H^2 G_4 + 24 H^2 X (G_{4,X} + X G_{4,XX}) \nonumber \\
&\quad - 12 H X \dot{\phi} G_{4,\phi X} - 6 H \dot{\phi} G_{4,\phi} + 2 H^3 X \dot{\phi} (5 G_{5,X} + 2X G_{5,XX}) \nonumber \\
&\quad - 6 H^2 X (3 G_{5,\phi} + 2X G_{5,\phi X}) = \rho_m
\end{align}
\begin{align}\label{9}
&G_2 - 2X \left(G_{3,\phi} + \ddot{\phi} G_{3,X} \right) + 2 G_4 (3H^2 + 2\dot{H}) - 12 H^2 X G_{4,X} - 4 H \dot{X}G_{4,X} \nonumber \\
&\quad -8\dot{H} X G_{4,XX}-8HX\dot{X}G_{4,XX} + 2 G_{4,\phi} (\ddot{\phi} + 2H\dot{\phi}) + 4X G_{4,\phi\phi} + 4X (\ddot{\phi} - 2H\dot{\phi}) G_{4,\phi X} \nonumber \\
&\quad + 4 H X \dot{\phi} \dot{G}_{5,\phi\phi} - 2X (2H^3 \dot{\phi} + 2H\dot{H} \dot{\phi} + 3H^2 \ddot{\phi}) G_{5,X} \nonumber \\
&\quad - 4 H^2 X^2 \ddot{G}_{5,XX} + 4 H X (\dot{X} - H X) G_{5,\phi X} + 2\left(\frac{d}{dt}(H X) + 3 H^2 X\right) = -p_m
\end{align}
Here, an overdot denotes differentiation with respect to cosmic time $t$ and the Hubble parameter is defined as $H = \frac{\dot{a}}{a}$, where $a(t)$ is the scale factor. In addition, differentiating the action (\ref{6}) by treating $\phi(t)$ as the variable gives the scalar field’s evolution equation in this cosmological background.
\begin{equation}\label{10}
\frac{1}{a^{3}}\frac{d(a^{3}J)}{dt}=P_{\phi},
\end{equation}
Here,
\begin{align}
J &= \dot{\phi} G_{2,X} + 6 H X G_{3,X} - 2 \dot{\phi} G_{3,\phi} + 6 H^2 \dot{\phi} (G_{4,X} + 2X G_{4,XX}) \\\nonumber
&\quad - 12 H X G_{4,\phi X} + 2 H^3 X (3 G_{5,X} + 2X G_{5,XX}) - 6 H^2 \dot{\phi} (G_{5,\phi} + X G_{5,\phi X}) \\
P_{\phi} &= G_{2,\phi} - 2X \left(G_{3,\phi\phi} + \ddot{\phi} G_{3,\phi X}\right) + 6 G_{4,\phi} (2 H^2 + \dot{H}) \\\nonumber
&\quad + 6 H G_{4,\phi X} (\dot{X} + 2 H X) - 6 H^2 X G_{5,\phi\phi} + 2 H^3 X \dot{\phi} G_{5,\phi X}\nonumber
\end{align}

We consider a specific class of scalar-torsion gravity theory, commonly referred to as the torsion Galileon model, formulated within the teleparallel gravity framework. The model is constructed by specifying the Galileon functions $G_i(\phi,X)$, where $i=2,3,4,5$, as follows:
\begin{equation}\label{13}
G_2(\phi, X)=X-V(\phi),\;\; G_3(\phi,X)=-F(\phi)X,\;\; G_4(\phi,X)=\frac{1}{2\kappa^2},\;\; G_5(\phi, X)=0.
\end{equation}

These functional forms lead to a modified set of cosmological field equations, which includes the Friedmann equations and the generalized Klein-Gordon equation for the scalar field $\phi$.
\begin{equation}\label{14}
3H^2= \kappa^2 \left( \frac{1}{2} \dot{\phi}^2 + V(\phi) - 3H \dot{\phi}^3 F(\phi) + \frac{1}{2} \dot{\phi}^4 F_{\phi} + \rho_m \right),
\end{equation}
\begin{equation}\label{15}
-2\dot{H}=\kappa^2 \left( \dot{\phi}^2 + \dot{\phi}^4 F_{\phi} + \ddot{\phi} \dot{\phi}^2 F(\phi) - 3H \dot{\phi}^3 F(\phi) + p_m \right).
\end{equation}
\begin{align}\label{16}
\ddot{\phi} + 3H \dot{\phi} + 2 \ddot{\phi} \dot{\phi}^2 F_{\phi} + \frac{1}{2} \dot{\phi}^4 F_{\phi\phi} - 3 \dot{H} \dot{\phi}^2 F(\phi) - 6H \ddot{\phi} \dot{\phi} F_{\phi} - 9H^2 \dot{\phi}^2 F(\phi) + V_{\phi} = 0.
\end{align}
The modified Friedmann equations obtained in equations (\ref{14}) and (\ref{15}) can be reformulated to resemble the conventional structure of standard cosmological equations.
\begin{equation}\label{17}
3H^2=\kappa^{2}(\rho_m+\rho_{de}),
\end{equation}
\begin{equation}\label{18}
-2\dot{H}=\kappa^2(\rho_m+\rho_{de}+p_{de}),
\end{equation}
where $\rho_{de}$ and $p_{de}$ denote the effective energy density and pressure arising from the scalar field sector. By comparing these expressions with the full modified Friedmann equations, one can isolate the scalar field contributions as:
\begin{equation}\label{19}
\rho_{de}=\frac{1}{2} \dot{\phi}^2 + V(\phi) - 3H \dot{\phi}^3 F(\phi) + \frac{1}{2} \dot{\phi}^4 F_{,\phi},
\end{equation}
\begin{equation}\label{20}
p_{de}=\frac{1}{2} \dot{\phi}^2 - V(\phi) + \dot{\phi}^2 \ddot{\phi} F(\phi) + \frac{1}{2} \dot{\phi}^4 F_{,\phi},
\end{equation}
These define the dark energy sector emerging from the torsion-coupled scalar field dynamics. From these, the equation-of-state (EoS) parameter for dark energy is given by:
\begin{equation}\label{21}
\omega_{de}=\frac{p_{de}}{\rho_{de}} =\frac{\frac{1}{2} \dot{\phi}^2 - V(\phi) + \dot{\phi}^2 \ddot{\phi} F(\phi) + \frac{1}{2} \dot{\phi}^4 F_{,\phi}
}{\frac{1}{2} \dot{\phi}^2 + V(\phi) - 3H \dot{\phi}^3 F(\phi) + \frac{1}{2} \dot{\phi}^4 F_{,\phi}},
\end{equation}
An important feature of Galileon cosmology is the dynamical flexibility of $\omega_{de}$, which can behave like quintessence, phantom, or even cross the phantom divide during evolution — depending on the functional form of $F(\phi)$. Additionally, the scalar field sector satisfies its own continuity equation, ensuring energy conservation:
\begin{equation}\label{22}
\dot{\rho}_{de}+3H(\rho_{de}+p_{de})=0,
\end{equation}
For the matter component, the energy density evolves as expected for pressureless dust:
\begin{equation}\label{23}
\dot{\rho}_{m}+3H\rho_{m}=0.
\end{equation}
\section{Approximate solutions via hubble parametrization in Galileon Cosmology}\label{sec3}
\hspace{0.5cm} In Galileon cosmology, the field equations derived from the variation of the action contain a combination of nonlinear terms, including the scalar field $\phi$, its derivatives (often up to second or third order), the function $F(\phi)$ and the potential $V(\phi)$. This complex structure makes the task of finding exact analytical solutions from the modified Friedmann equations highly non-trivial. The system becomes particularly intricate due to the coupling between the Galileon field and the geometric sector of gravity. Rather than solving the system analytically, it is typical to proceed by postulating a particular behavior for the Hubble parameter $H(z)$, which enables further exploration of the model. This strategy reduces the number of unknowns and provides a way to extract physically meaningful quantities, such as the evolution of the scalar field and the effective equation of state. To proceed with our investigation, we make use of the Hubble parameter formulation presented in \cite{Yu24}:
\begin{equation}\label{24}
H^{2}(z)=H_{0}^{2}\left[ \Omega_{m0}(1+z)^{3}+\Omega_{de}(z)\right],
\end{equation}
where $\Omega_{m0}$ denotes the present-day density parameter for non-relativistic matter (including baryons and dark matter) and $\Omega_{de}(z)$ accounts for the effective energy contribution from the dark energy. Unlike the standard dark energy component, $\Omega_{de}(z)$ originates from the scalar field dynamics governed by the DE. The radiation term $\Omega_{r0}(1+z)^4$ is omitted in our analysis, given that its influence is minimal in the late Universe, where matter and dark energy prevail. This assumption is well-justified because radiation dominates only at very early epochs (high redshift $z\gg 10^{4}$), and has minimal impact on the background dynamics at low and intermediate redshifts relevant to current observational constraints.

For a general scalar field with a time-dependent equation of state $\omega_{de}(z)=\frac{p_{de}}{\rho_{de}}$, the DE density parameter evolves as:
\begin{equation}\label{25}
\Omega_{DE}(z)=\Omega_{DE0}\exp\left[3\int_0^z \frac{1+\omega_{DE}(z')}{1+z'}\,dz'\right],
\end{equation}
Given the assumption of a flat Universe, the gravitational component satisfies $\Omega_{DE0}=1-\Omega_{m0}$. To characterize the evolution of the equation of state parameter for the DE, we utilize the widely used Chevallier–Polarski–Linder (CPL) parametrization \cite{Ch01,EV03}:
\begin{equation}\label{26}
\omega_{DE}(z)=\omega_{0}+\omega_{a}\frac{z}{1+z},
\end{equation}
where $\omega_{0}$ represents the present-day value of the dark energy equation of state and $\omega_{a}$ quantifies its evolution with redshift. This parametrization is widely used in the literature due to its ability to effectively capture the dynamics of dark energy in both low- and high-redshift regimes. At present time ($z=0$), we have $\omega_{DE}(0)=\omega_{0}$, while at early times ($z \gg 1$), the equation of state approaches $\omega_{DE}(z)\rightarrow\omega_{0}+\omega_{a}$. Substituting this form into the equation (\ref{26}) and integrate, we obtain the Hubble parameter evolution as:
\begin{equation}\label{27}
H(z)=H_{0}\left[\Omega_{m0}(1+z)^3+(1 - \Omega_{m0})(1+z)^{3(1+\omega_0+\omega_a)} \exp\left(-\frac{3 \omega_a z}{1+z} \right) \right].
\end{equation}
Here, $H_{0}$ is the current Hubble constant. We proceed by employing cosmological observations to estimate the values of these parameters.
\subsection{Parameter estimation using observational datasets}\label{sec3.1}
\hspace{0.5cm} In the preceding section, we presented a parameterized expression for the Hubble parameter within the context of Galileon cosmology. Constraining the model's parameters with recent cosmological data allows us to assess its observational relevance. Our analysis incorporates a set of $46$ Hubble parameter data points, complemented by the latest BAO observations from DESI DR2 and a comprehensive sample of $1701$ supernovae of Type Ia. The model includes four free parameters: the Hubble constant $H_{0}$, $\Omega_{m0}$ for the current matter density and $\omega_{0}$ and $\omega_{a}$, which describe the dark energy sector's EoS. Markov Chain Monte Carlo (MCMC) simulations are conducted with the \texttt{emcee} \cite{Mackey13} package to derive best-fit parameter values, which efficiently explores multi-dimensional parameter spaces. The parameter ranges explored in the MCMC analysis are: $60<H_{0}<80$, $0<\Omega_{m0}<1$, $-1<\omega_{0}<1$ and $-1<\omega_a<2$. We employ $100$ walkers and $1500$ iterations per chain to ensure sufficient convergence and sampling of the posterior distributions. The total likelihood is constructed by combining contributions from all three datasets, based on the standard Gaussian form: $\mathcal{L}\propto e^{-\frac{\chi^{2}}{2}}$, where $\chi^{2}$ measures the discrepancy between the theoretical model and observational data for each dataset. By optimizing the total likelihood function, we derive reliable and comprehensive constraints on the model parameters. 
\subsubsection{Hubble data analysis}\label{sec3.1.1}
\hspace{0.5cm} The Hubble parameter provides a direct and model-independent tracer of the Universe's expansion history. Our study utilizes $46$ independent $H(z)$ data points distributed across the redshift interval $z\in[0, 2.36]$. These measurements are obtained through the cosmic chronometer approach and the differential age method of galaxies. The data compilation draws from multiple observational studies, which includes the recent surveys in \cite{D10,M12,Sam24}, ensuring comprehensive coverage from low to high redshift regimes. A key strength of the cosmic chronometer method is its model-independence, based on the fundamental relation: $H(z)=-(1+z)^{-1}\frac{dz}{dt}$, where $\frac{dz}{dt}$ is obtained by estimating differential ages of passively evolving galaxies. To statistically constrain the cosmological parameters, we construct the chi-squared function corresponding to the Hubble data as:
\begin{equation}\label{28}
\chi^{2}_{H} = \sum_{i=1}^{46}\frac{\left[ H_{th}(z_i) - H_{obs}(z_i) \right]^2}{\sigma_H(z_i)^2},
\end{equation}
$H_{obs}(z_i)$ and $\sigma_H(z_i)$ correspond to the measured Hubble parameter and its associated uncertainty at $z_i$, and $H_{th}(z_i)$ symbolizes the theoretical prediction from the model.
\subsubsection{Baryon Acoustic Oscillations from DESI DR2}\label{sec3.1.2}
\hspace{0.5cm} In this analysis, we employ the most up-to-date Baryon Acoustic Oscillation (BAO) data from the Dark Energy Spectroscopic Instrument (DESI) survey’s Data Release 2 (DR2) \cite{Karim2025}. These observations span a diverse set of cosmological tracers: the Bright Galaxy Sample (BGS), three tiers of Luminous Red Galaxies (LRG1–3), two classifications of Emission Line Galaxies (ELG1–2), Quasars (QSO), and the Lyman-$\alpha$ forest. The BAO signal corresponds to the comoving scale of the sound horizon at the baryon drag epoch (approximately $z_d \sim 1060$), which is characterized by the integral expression:
\begin{equation}\label{29}
r_d = \int_{z_d}^{\infty} \frac{c_s(z)}{H(z)} \, dz,
\end{equation}
where $c_s(z)$ symbolizes the sound speed in the tightly coupled photon-baryon fluid. According to the $\Lambda$CDM framework, $r_d = 147.09 \pm 0.2$ Mpc is the commonly adopted value for the sound horizon at the drag epoch, as derived from Planck 2018 \cite{Planck2018}. Our study relaxes the fixed value of $r_d$, instead treating it as a free parameter to investigate possible extensions beyond the standard cosmology \cite{DESI24,AG25,AGA25,Pogo20,Vag23}. Three key cosmological distance measures are derived to facilitate a comparison between theoretical predictions and actual BAO observations.

\vspace{0.2cm}

\noindent
\textbf{Hubble distance $(D_H(z))$:}
\begin{equation}\label{30}
D_H(z) = \frac{c}{H(z)},
\end{equation}
\noindent
\textbf{Comoving-coordinate angular distance $(D_M(z))$:}
\begin{equation}\label{31}
D_M(z) = c \int_0^z \frac{dz'}{H(z')},
\end{equation}
\noindent
\textbf{Effective distance measure averaged over volume $D_V(z)$:}
\begin{equation}\label{32}
D_V(z) = \left[ z D_M^2(z) D_H(z) \right]^{1/3}.
\end{equation}
These distance measures are often made dimensionless through division by the sound horizon scale, which lead to the ratios $\frac{D_H}{r_d}$, $\frac{D_M}{r_d}$, $\frac{D_V}{r_d}$ and $\frac{D_M}{D_H}$. Chi-squared function assesses theory and BAO data consistency as:
\begin{equation}\label{33}
\chi^2_{BAO} = \Delta D^T \, C^{-1} \, \Delta D,
\end{equation}
where $\Delta D = D_{obs} - D_{th}$ is the deviation of observed from theoretical distances and $C^{-1}$ being the inverse of the BAO data covariance matrix.
\subsubsection{Pantheon+ supernovae observations}\label{sec3.1.3}
\hspace{0.5cm} Type Ia Supernovae (SNe Ia) are valuable for understanding the cosmos because of their well-known intrinsic brightness. Over the years, the number of observed SNe Ia has significantly increased, leading to more robust constraints on cosmological models. In this analysis, we utilize the Pantheon+ compilation—the most comprehensive collection of SNe Ia to date. This dataset consolidates observations from multiple surveys, including the Sloan Digital Sky Survey (SDSS), the Supernova Legacy Survey (SNLS), the Hubble Space Telescope (HST) and others, resulting in a total of $1701$ spectroscopically confirmed supernovae gathered from $18$ distinct datasets \cite{Riess22,Samaddar25}. The Pantheon+ sample spans a wide redshift interval, ranging from $z = 0.0012$ to $z = 2.2614$, with a denser sampling at low redshifts. Observed distance moduli $\mu_{obs}$ are contrasted with theoretical expectations $\mu_{th}$ to determine cosmological parameters. One gets the distance modulus theoretically through:
\begin{equation}\label{34}
\mu_{th}(z) = 5 \log_{10} \left[ \frac{D_L(z)}{Mpc} \right]+25,
\end{equation}
Here, $D_L(z)$ denotes the luminosity distance, characterized by:
\begin{equation}\label{35}
D_L(z) = c (1 + z) \int_0^z \frac{d\tilde{z}}{H(\tilde{z}; H_{0},\Omega_{0},\omega_{0},\omega_{a})},
\end{equation}
The fit of the theoretical model to observational data is assessed with a chi-squared ($\chi^2$) statistical test, formulated as:
\begin{equation}\label{36}
\chi^2_{SN}=\sum_{i,j=1}^{1701} \Delta \mu_i \left[C_{SN}^{-1}\right]_{ij} \Delta \mu_j,
\end{equation}
where 
\begin{equation}\label{37}
\Delta \mu_i = \mu_{th,i} - \mu_{obs,i},
\end{equation}
and statistical and systematic errors are captured in the covariance matrix $C_{SN}$. 
\subsubsection{Combined statistical analysis incorporating multiple cosmological probes}\label{sec3.1.4}
\begin{figure}[hbt!]
  \centering
  \includegraphics[scale=0.38]{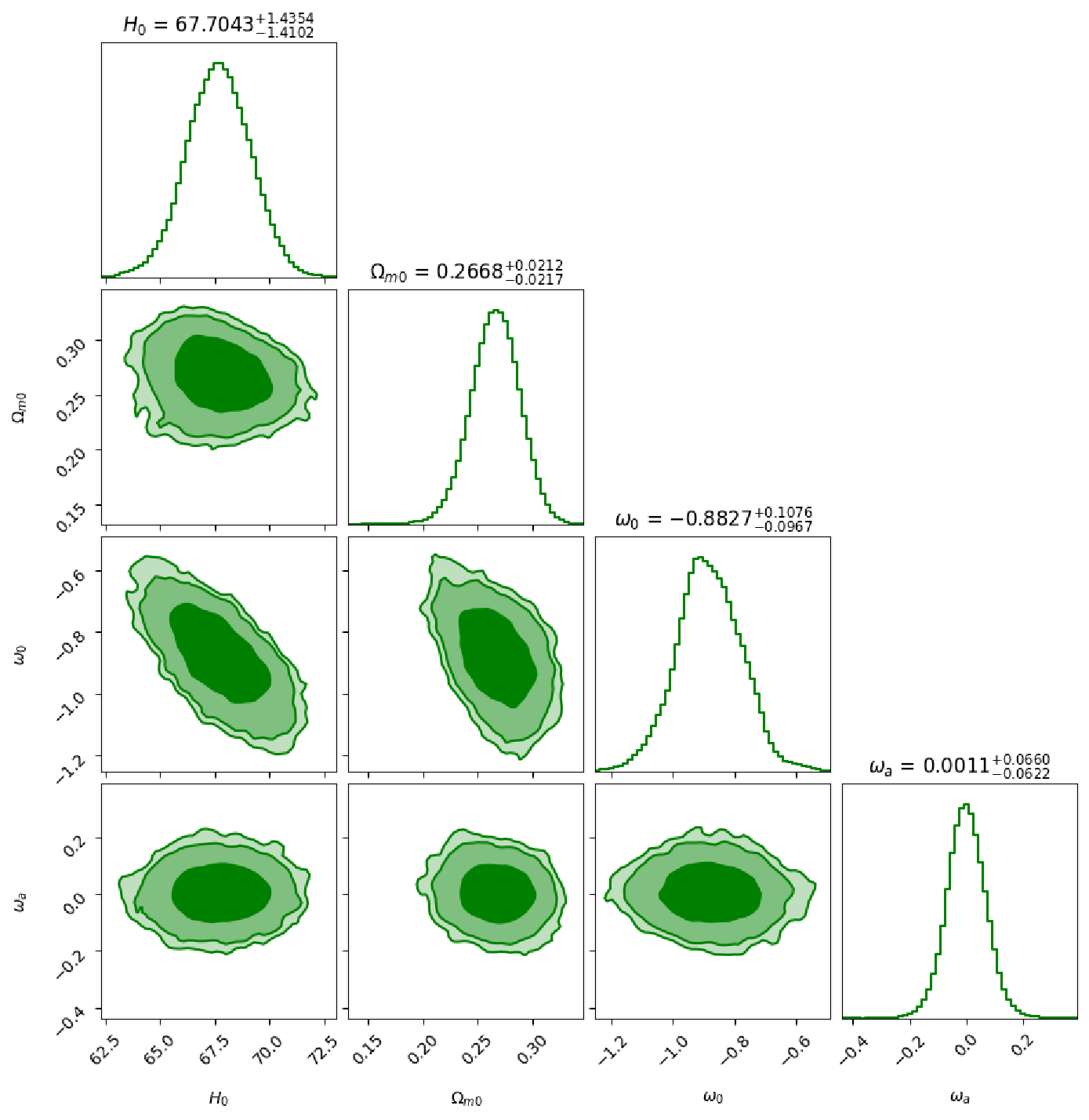}
  \caption{Two-dimensional joint confidence regions for the parameters $(H_{0},\Omega_{0},\omega_{0},\omega_{a})$ at $1\sigma$, $2\sigma$ and $3\sigma$ confidence levels, obtained from the combined investigation of Hubble, Pantheon and DESI DR2 BAO datasets.}\label{fig:f1}
\end{figure}

\begin{figure}[hbt!]
  \centering
  \includegraphics[scale=0.43]{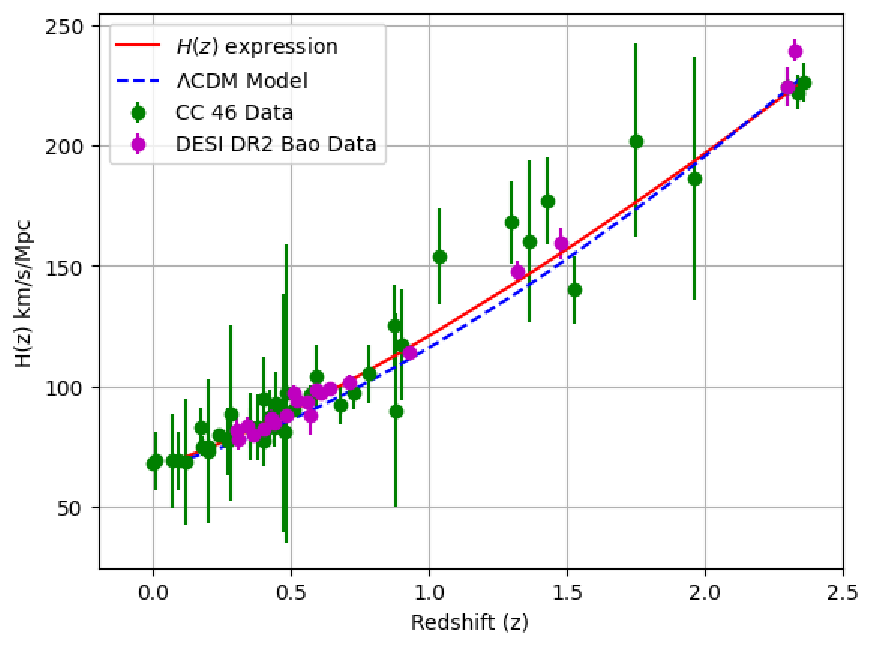}~~
  \includegraphics[scale=0.33]{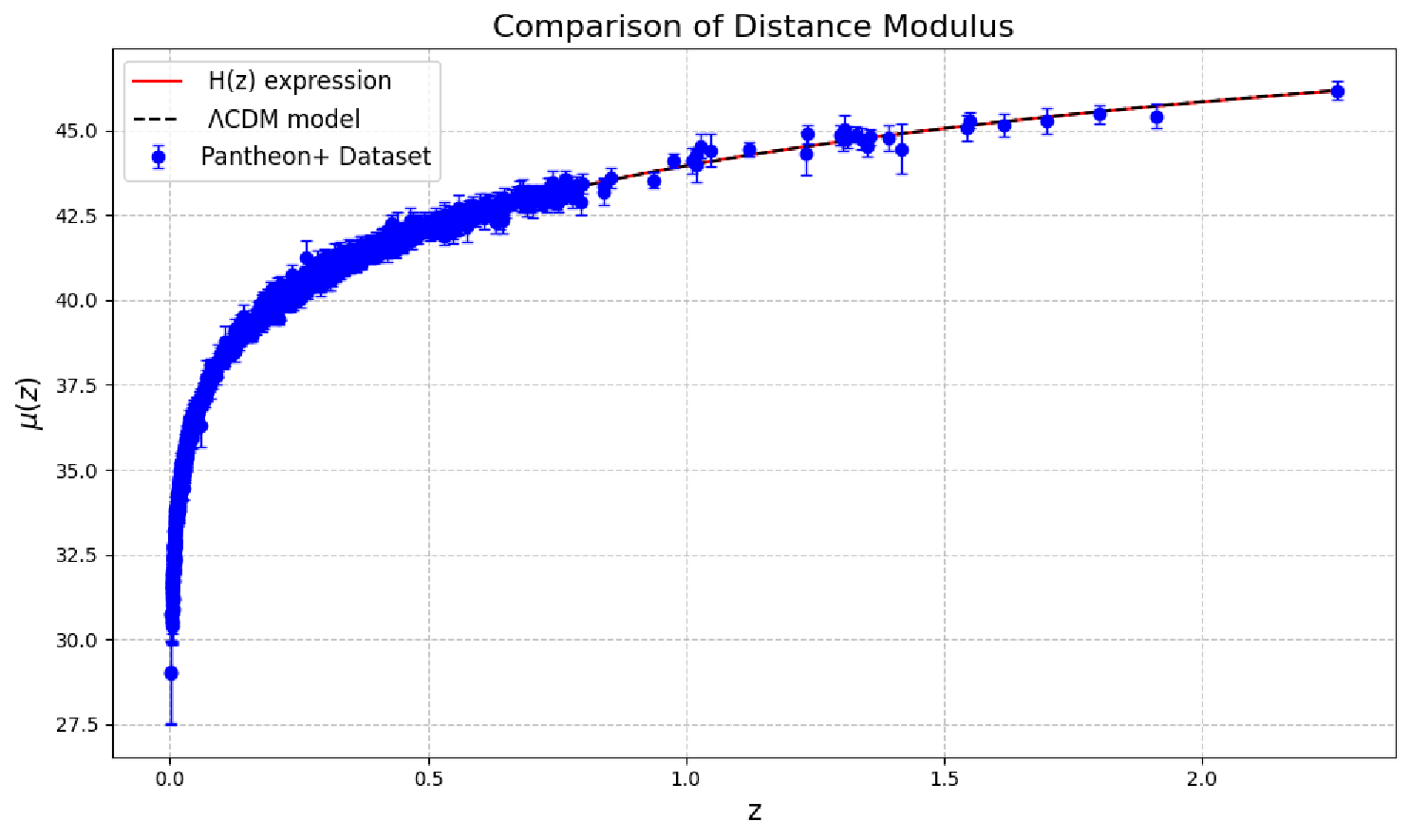}
  \caption{Error bar comparison for derived $H(z)$ model (red) and the $\Lambda$CDM model (dashed line) using observational datasets. Left panel: Hubble + DESI DR2 BAO; Right panel: Pantheon+ dataset.}\label{fig:f2}
\end{figure}

\hspace{0.5cm} To derive tighter constraints on the cosmological parameters, we carry out a unified analysis that integrates various observational datasets, including the Pantheon+ Type Ia Supernovae sample, BAO measurements and $H(z)$ data. The cumulative likelihood function is formulated by aggregating the individual chi-square contributions from each dataset. Specifically, the total chi-square used in the parameter inference is expressed as:
\begin{equation}\label{38}
\chi^2_{total}=\chi^2_{SN}+\chi^2_{BAO}+\chi^2_{H},
\end{equation}
where each term represents the chi-square derived from the corresponding SNe Ia, BAO and Hubble datasets. The most accurate estimates of the model parameters are obtained by reducing the total chi-square function, which integrates the constraints from diverse cosmological observations. Figures \ref{fig:f1} and \ref{fig:f2} provide a visual summary of the parameter estimation results obtained from our analysis. Figure \ref{fig:f1} presents the contour plots at $1\sigma$ $(68.3\%)$, $2\sigma$ $(95.4\%)$ and $3\sigma$ $(99.7\%)$ confidence levels for the cosmological model parameters. These contours illustrate the statistically allowed regions in parameter space and reveal correlations among the parameters constrained by the joint datasets. A plot of the $H(z)$ against redshift appears in Figure \ref{fig:f2}, where the model's theoretical predictions are plotted alongside the observational data with error bars. The close alignment between the model curve and the observational uncertainties highlights the consistency and reliability of the proposed model.

From the joint likelihood analysis using the Hubble, DESI DR2 BAO and Pantheon+ datasets, we obtain the following best-fit values and $1\sigma$ uncertainties for the model parameters: $H_{0}= 67.7043^{+1.4354}_{-1.4102}$ \text{km/s/Mpc}, $\Omega_{m0}= 0.2668^{+0.0212}_{-0.0217}$, $\omega_0= -0.8827^{+0.1076}_{-0.0967}$ and $\omega_a= 0.0011^{+0.0660}_{-0.0622}$. These results are in excellent agreement with recent cosmological measurements. The obtained Hubble constant $H_0$ is consistent with the value reported by the Planck 2018 findings within the $\Lambda$CDM model, which gives $H_0 = 67.4 \pm 0.5$ km/s/Mpc \cite{Planck2018}. The value of $\Omega_{m0}$ also lies within the range reported by both Planck \cite{Planck2018} and DESI DR 1 results \cite{DESIY1}. The EoS $\omega_0$ and $\omega_a$ are in agreement with current constraints from combined datasets (CMB + BAO + SNe Ia), where the equation of state typically satisfies $\omega_0 \in [-1.1, -0.9]$ and $\omega_a \approx 0$ \cite{Scolnic22}. Our model shows strong consistency with the latest cosmological observations. The values do not require any exotic deviations from the $\Lambda$CDM framework but instead demonstrate that our parametrized $H(z)$ model is capable of accurately reproducing the background expansion evolution of the Universe. Furthermore, the combination of datasets (Hubble + DESI and Pantheon+) enhances the robustness of the parameter estimation, confirming the consistency and viability of our model.
\subsubsection{Model selection criteria: AIC and BIC analysis}\label{sec3.1.5}
\hspace{0.5cm} The Akaike and Bayesian Information Criteria (AIC and BIC) are employed to statistically compare our cosmological model with the conventional $\Lambda$CDM framework. These criteria penalize excessive free parameters to prevent overfitting and balance model complexity with goodness of fit. The AIC is given by the formula \cite{Liddle04,Ness13,Akaike74}:
\begin{equation}\label{39}
AIC: \quad \chi^2_{\min} + 2k,
\end{equation} 
where $\chi^2_{\min}$ is the smallest chi-square value obtained during the fitting process, and $k$ stands for the number of adjustable parameters in the model. Although the AIC accounts for how well a model fits the data, it also includes a correction term that grows with the number of free parameters. A model with a lower AIC is generally preferred. To judge the relative effectiveness of each model, we compute the difference:
\begin{equation}\label{40}
\Delta AIC=|AIC_{model}-AIC_{\Lambda CDM}|,
\end{equation}
The BIC is more conservative than AIC, which applies a stronger penalty for model complexity. It is expressed as \cite{Sch78}:
\begin{equation}\label{41}
BIC: \quad \chi^2_{\min} + k \ln N,
\end{equation}
where $N$ stands for the count of data points involved in the study. The corresponding difference is:
\begin{equation}\label{42}
\Delta BIC=|BIC_{model}-BIC_{\Lambda CDM}|,
\end{equation}
According to widely accepted guidelines, if $0\leq|\Delta AIC|< 2$, there is strong support for both models, which indicates that they are statistically equivalent. When $2 \leq |\Delta AIC|<4$, there is moderate evidence against the model with the higher AIC value. Finally, if $|\Delta AIC|\geq4$, it indicates strong evidence against the model with the higher AIC. For the Bayesian Information Criterion (BIC), if $0 \leq |\Delta BIC| < 2$, there is only weak evidence against the model with the higher BIC value. A difference in the range $2 \leq |\Delta BIC| < 6$ indicates moderate evidence, while $6 \leq |\Delta BIC| < 10$ corresponds to strong evidence. Finally, if $|\Delta BIC| \geq 10$, it implies very strong evidence against the model with the higher BIC.

For our cosmological model fitted using the joint dataset (Hubble + DESI DR2 BAO + Pantheon+), we obtain a minimum chi-square value of $\chi^2_{\min}=158.7$, with $k=4$ free parameters and $N =1768$ data points. The AIC is then calculated as $AIC= 163.24$, which yields a difference of $\Delta AIC=1.46$. Similarly, the BIC is computed as $BIC=185.15$, which results in $\Delta BIC=11.5$. The AIC result ($\Delta AIC=1.46$) falls well within the range of strong statistical support, which suggests that our model is competitive and comparable to the $\Lambda$CDM model. The BIC value ($\Delta BIC=11.5$), however, exceeds the threshold of $10$, which indicate very strong preference for $\Lambda$CDM under BIC's more stringent complexity penalty.

These results imply that while our model achieves a very good fit to the data and is favored under AIC, the increased number of parameters slightly reduces its appeal under the BIC criterion. Nevertheless, the consistency across multiple datasets and the statistical closeness in AIC confirm the viability of our $H(z)$-based dynamical dark energy model as a plausible alternative to the standard cosmological scenario.
\section{Cosmological parameters and evolution}\label{sec4}
\hspace{0.5cm} In this section, we examine essential cosmological quantities that characterize the Universe's expansion behavior. They provide valuable information about the deceleration-to-acceleration transition and serve as tools to test the observational consistency of our model.
\subsection{Deceleration parameter}\label{sec4.1}
\hspace{0.5cm} A fundamental parameter that describes the Universe's expansion dynamics is the deceleration parameter $q(z)$, which indicates whether the cosmic expansion is accelerating or decelerating. When the deceleration parameter $q(z)$ takes on a negative value, it reflects an accelerating Universe, whereas a positive value corresponds to a decelerating expansion. Based on the Hubble parameter obtained from our model, using equation (\ref{27}) the deceleration parameter can be explicitly expressed as:
\begin{eqnarray}\label{43}
q(z)&=&-1+\frac{1}{2\bigg[\Omega_{m0}(1+z)^3 + (1-\Omega_{m0})(1+z)^{3(1+\omega_0 + \omega_a)} \exp\left(-\frac{3 \omega_a z}{1+z}\right)\bigg]}\times\\\nonumber
&&3\bigg[\Omega_{m0}(1+z)^3 + (1 - \Omega_{m0})(1+\omega_0 + \omega_a)(1+z)^{3(1+\omega_0 + \omega_a)} \exp\left(-\frac{3 \omega_a z}{1+z}\right)\\\nonumber
&&-(1-\Omega_{m0})(1+\omega_0 + \omega_a)(1+z)^{2 + 3\omega_0 + 3\omega_a} \omega_a \exp\left(-\frac{3 \omega_a z}{1+z}\right)\bigg].
\end{eqnarray}
\begin{figure}[hbt!]
  \centering
  \includegraphics[scale=0.4]{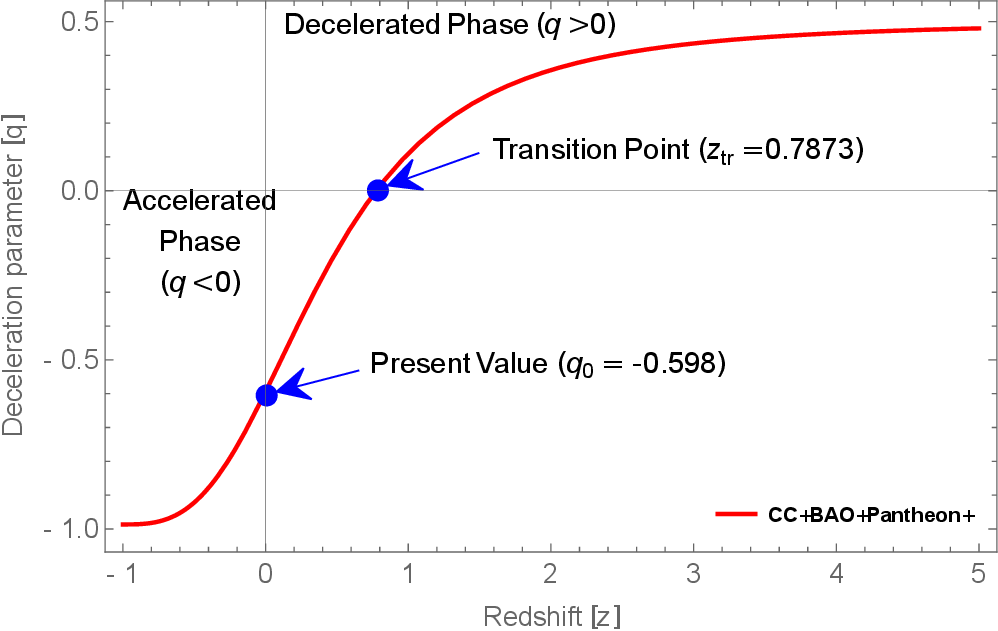}
  \caption{Behavior of the deceleration parameter $q(z)$ across redshift.}\label{fig:f3}
\end{figure}

Figure \ref{fig:f3} presents the evolution of the deceleration parameter $q(z)$ as a function of redshift. From the plot, it is evident that at higher redshifts (corresponding to the early Universe), the expansion is decelerating, with $q\approx 0.474$. As the Universe evolves, it transitions from deceleration to acceleration at the redshift $z_{tr}=0.7873$, which marks the transition point. At late times, the deceleration parameter tends toward $q=-1$, replicating the behavior of the standard $\Lambda$CDM model. The present-day value of the deceleration parameter is $q_{0}=-0.598$, which confirms the currently observed accelerated expansion of the Universe and agrees well with contemporary observational constraints \cite{ASamaddar25}.
\subsection{Energy density and pressure in Galileon Cosmology}\label{sec4.2}
\hspace{0.5cm} To investigate the behavior of the energy density and pressure in Galileon cosmology, we make use of the modified Friedmann equations, which involve contributions from the scalar field $\phi$, the potential $V(\phi)$ and the Galileon coupling function $F(\phi)$. Since these functions are not uniquely fixed by theory, we must assume plausible forms motivated by prior studies in the literature. In this analysis, we adopt the following functional forms:
\begin{equation}\label{44}
\phi(z)=\phi_{0}(1+z)^{-m}, \hspace{0.5cm} V(\phi)=V_{0}\phi^{n}, \hspace{0.5cm} F(\phi)=F_{0}e^{-\lambda l\phi(t)},
\end{equation}
where $\phi_{0}$, $V_{0}$, $F_{0}$, $\lambda$ and $l$ are constants. These forms are commonly used in scalar-tensor and Galileon cosmology frameworks \cite{Felice11,Gann10} and facilitate the direct comparison with observationally motivated Hubble expansion histories. Substituting these expressions into the Friedmann equations (\ref{19}) and (\ref{20}) and utilizing our model for $H(z)$ from equation (\ref{27}), we derive the corresponding expressions for the dark energy density $\rho_{de}(z)$ and pressure $p_{de}(z)$ of the Galileon field as follows:
\begin{eqnarray}\label{45}
\rho_{de}&=&\frac{m^{2}\phi_{0}^{2}(1+z)^{-2m}H_{0}^{2}}{2}\bigg[\Omega_{m0}(1+z)^3 + (1-\Omega_{m0})(1+z)^{3(1+\omega_0 + \omega_a)} \exp\left(-\frac{3 \omega_a z}{1+z}\right)\bigg]\\\nonumber
&&+V_{0}\bigg[\phi_{0}(1+z)^{-m}\bigg]^{n}-3m^{3}\phi_{0}^{3}(1+z)^{-3m}H_{0}^{3}\bigg[\Omega_{m0}(1+z)^3 + (1-\Omega_{m0})(1+z)^{3(1+\omega_0 + \omega_a)}\\\nonumber
&&\exp\left(-\frac{3 \omega_a z}{1+z}\right)\bigg]^{\frac{3}{2}}-\frac{m^{4}\phi_{0}^{4}\lambda lF_{0}H_{0}^{4}e^{-\lambda l\phi_{0}(1+z)^{-m}}}{2}\bigg[\Omega_{m0}(1+z)^3 +(1-\Omega_{m0})(1+z)^{3(1+\omega_0 + \omega_a)}\\\nonumber
&&\exp\left(-\frac{3 \omega_a z}{1+z}\right)\bigg]^{2},
\end{eqnarray}
\begin{eqnarray}\label{46}
p_{de}&=& \frac{m^{2}\phi_{0}^{2}(1+z)^{-2m}H_{0}^{2}}{2}\bigg[\Omega_{m0}(1+z)^3 + (1-\Omega_{m0})(1+z)^{3(1+\omega_0 + \omega_a)} \exp\left(-\frac{3 \omega_a z}{1+z}\right)\bigg]\\\nonumber
&&-V_{0}\bigg[\phi_{0}(1+z)^{-m}\bigg]^{n}-m^{3}\phi_{0}^{3}(1+z)^{-2m}H^{3}(z)F_{0}e^{-\lambda l\phi_{0}(1+z)^{-m}}\left[(1+z)^{-m+1}\frac{dH}{dz}-mH(z)(1+z)^{-m}\right]\\\nonumber
&&-\frac{m^{4}\phi_{0}^{4}\lambda lF_{0}H^{4}(z)(1+z)^{-4m}e^{-\lambda l\phi_{0}(1+z)^{-m}}}{2}.
\end{eqnarray}
\begin{figure}[hbt!]
  \centering
  \includegraphics[scale=0.4]{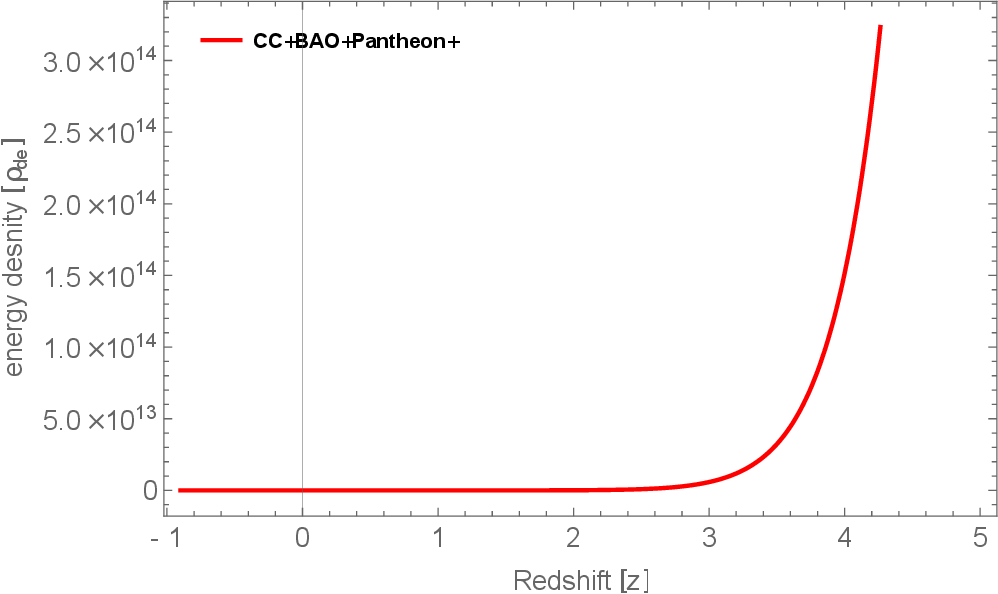}~~~
  \includegraphics[scale=0.4]{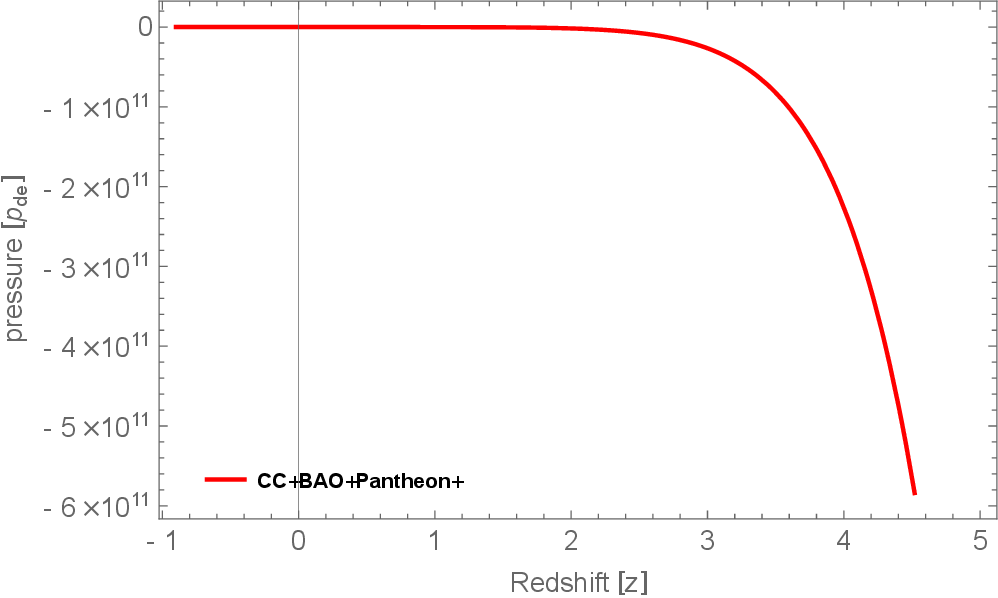}
  \caption{Evolution of energy density $\rho_{de}(z)$ and pressure $p_{de}(z)$ of the Galileon field for $\phi_{0}=1$, $V_{0}=0.05$, $F_{0}=0.01$ and $\lambda=-0.001$.}\label{fig:f4}
\end{figure}

Figure \ref{fig:f4} illustrates the evolution of the energy density and pressure of the Galileon field as functions of redshift. The plot shows that $\rho_{de}(z)$ remains positive throughout cosmic history, ensuring a physically viable energy content. Simultaneously, $p_{de}(z)$ remains negative for all redshifts, which supports the accelerated expansion of the Universe and is consistent with dark energy-like behavior.
\subsection{Equation of state parameter}\label{sec4.3}
\begin{figure}[h!]
  \centering
  \includegraphics[scale=0.42]{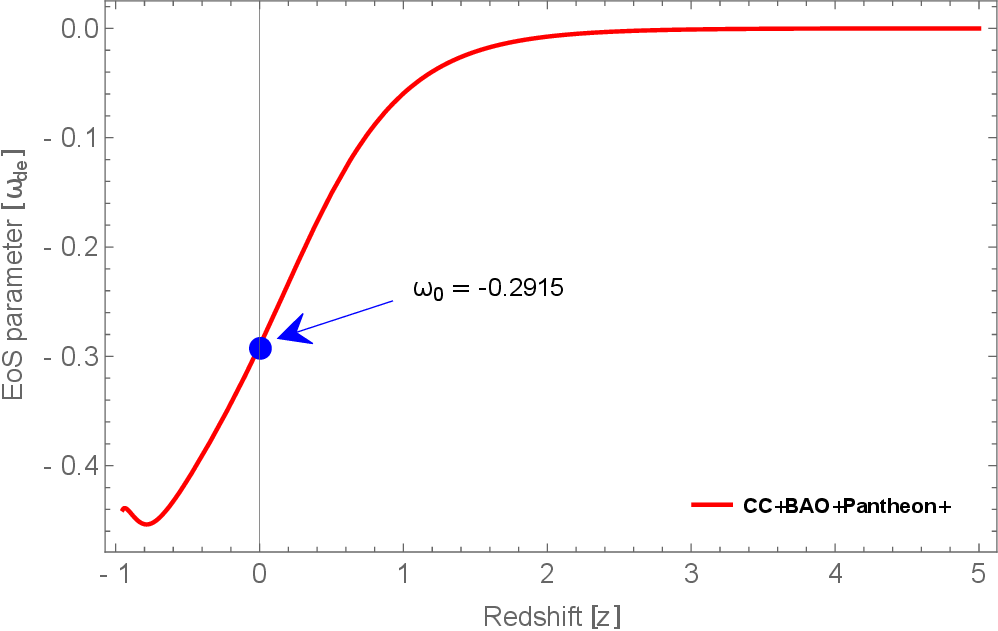}
  \caption{Evolution of the $\omega_{de}(z)$ as a function of redshift.}\label{fig:f5}
\end{figure}

\hspace{0.5cm} Using Equation (\ref{21}) along with the expressions for $\rho_{de}(z)$ and $p_{de}(z)$ given by Equations (\ref{27}) and (\ref{44}), we numerically compute the behavior of $\omega_{de}(z)$. The resulting plot is shown in Figure \ref{fig:f5}.

From Figure \ref{fig:f5}, we observe that the EoS parameter starts at $\omega(z)\approx0.00157$ at high redshift, indicating behavior close to a pressureless matter-like regime. As the redshift decreases, $\omega(z)$ declines and enters the dark energy regime. Eventually, the curve approaches the cosmological constant value $\omega=-1$, consistent with the $\Lambda$CDM model. The present-day value of the EoS parameter is $\omega_0=-0.2915$, which suggests a mildly dynamic dark energy component rather than a pure cosmological constant. This behavior reflects a viable and observationally consistent mechanism for late-time cosmic acceleration within the framework of Galileon cosmology.
\subsection{Energy conditions}\label{sec4.4}
\hspace{0.5cm} Energy conditions serve as fundamental criteria in gravitational physics to evaluate the physical plausibility of matter content and spacetime configurations. They impose restrictions on the energy-momentum tensor and are instrumental in analyzing the properties of cosmic fluids within the framework of general relativity and its extensions. In the context of cosmology, the standard energy conditions are defined as follows: Null Energy Condition (NEC) requires $\rho_{de}+p_{de}\geq 0$, which ensures that energy density as measured by any null observer is non-negative. The Strong Energy Condition (SEC) is satisfied when $\rho_{de}+3p_{de}\geq 0$, which is related to the attractive nature of gravity in general relativity. Finally, the Dominant Energy Condition (DEC) stipulates $\rho_{de}-p_{de}\geq 0$, which ensures that the energy flux is non-spacelike and the energy density dominates over pressure. These conditions are particularly relevant for understanding the acceleration or deceleration of the Universe and for testing the consistency of modified gravity models. Using our previously derived expressions (\ref{45}) and (\ref{46}) for the energy density $\rho_{de}(z)$ and pressure $p_{de}(z)$, the energy conditions can be expressed as:
\begin{equation}\label{47}
\rho_{de}+p_{de}=\dot{\phi}^{2}+\dot{\phi}^{2}F(\phi)\left(\ddot{\phi}-3\dot{\phi} H\right)+\dot{\phi}^{4}F_{,\phi},
\end{equation}
\begin{equation}\label{48}
\rho_{de}-p_{de}=2V(\phi)-\dot{\phi}^{2}F(\phi)\left(\ddot{\phi}+3\dot{\phi} H\right),
\end{equation}
\begin{equation}\label{49}
\rho_{de}+3p_{de}=2\dot{\phi}^{2}-2V(\phi)-3\dot{\phi}^{2}F(\phi)\left(\dot{\phi}H-\ddot{\phi}\right)+2\dot{\phi}^{4}F_{,\phi}.
\end{equation}
\begin{figure}[h!]
  \centering
  \includegraphics[scale=0.42]{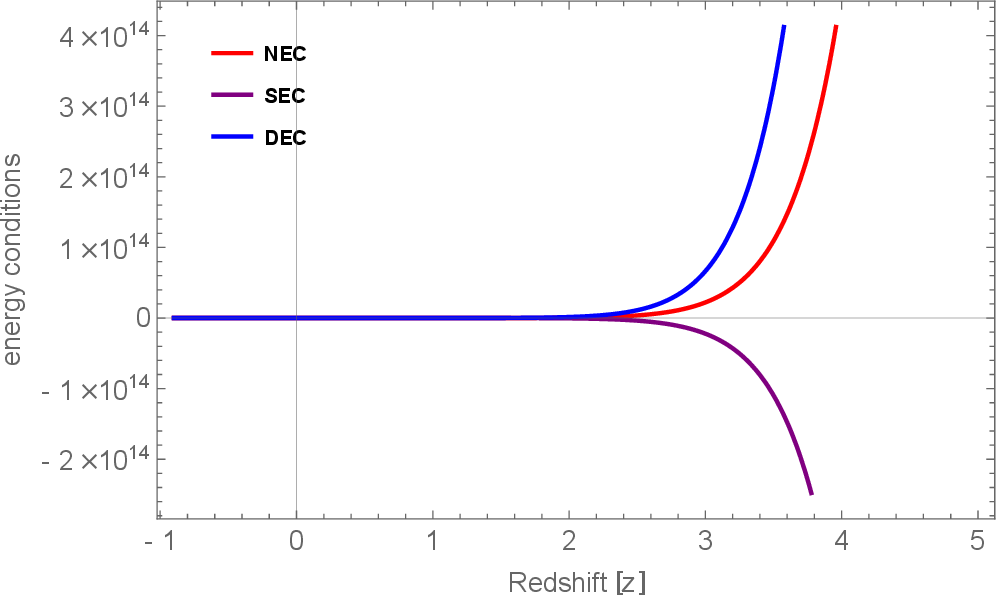}
  \caption{Evolution of the energy conditions with redshift.}\label{fig:f6}
\end{figure}

Figure \ref{fig:f6} illustrates the evolution of these energy conditions as functions of redshift. From Figure \ref{fig:f6}, it is evident that the NEC and the DEC hold at all epochs, which confirms that the energy flux remains causal and the energy density consistently exceeds the pressure, as expected for physically acceptable matter content. In contrast, the SEC is violated over the entire redshift range, a behavior characteristic of dark energy scenarios. This persistent violation indicates the presence of repulsive gravitational effects responsible for the observed accelerated expansion of the Universe.
\section{Statefinder diagnostics}\label{sec5}
\hspace{0.5cm} The statefinder diagnostic serves as an effective geometrical method for distinguishing between different dark energy models beyond the conventional $\Lambda$CDM framework. By incorporating higher-order derivatives of the scale factor, it offers a more refined perspective on the Universe’s expansion history. The statefinder pair $(r, s)$ is defined through the following expressions:
\begin{equation}\label{50}
r = \frac{\dddot{a}}{aH^3}, \quad s = \frac{r - 1}{3\left(q - \frac{1}{2}\right)},
\end{equation}
where $r$ captures the jerk (third derivative of the scale factor) and $s$ helps in distinguishing various cosmological scenarios based on their dynamic behavior. These parameters serve as a diagnostic pair for comparing the evolutionary behaviors of different cosmological models within the $(r, s)$ plane. The standard $\Lambda$CDM model is represented by the fixed point $(r,s)=(1,0)$. Quintessence and phantom dark energy models generally occupy separate regions in this plane. The path traced by a given model in the $(r, s)$ space provides insight into its alignment with or departure from the standard cosmological scenario.
\begin{figure}[hbt!]
  \centering
  \includegraphics[scale=0.44]{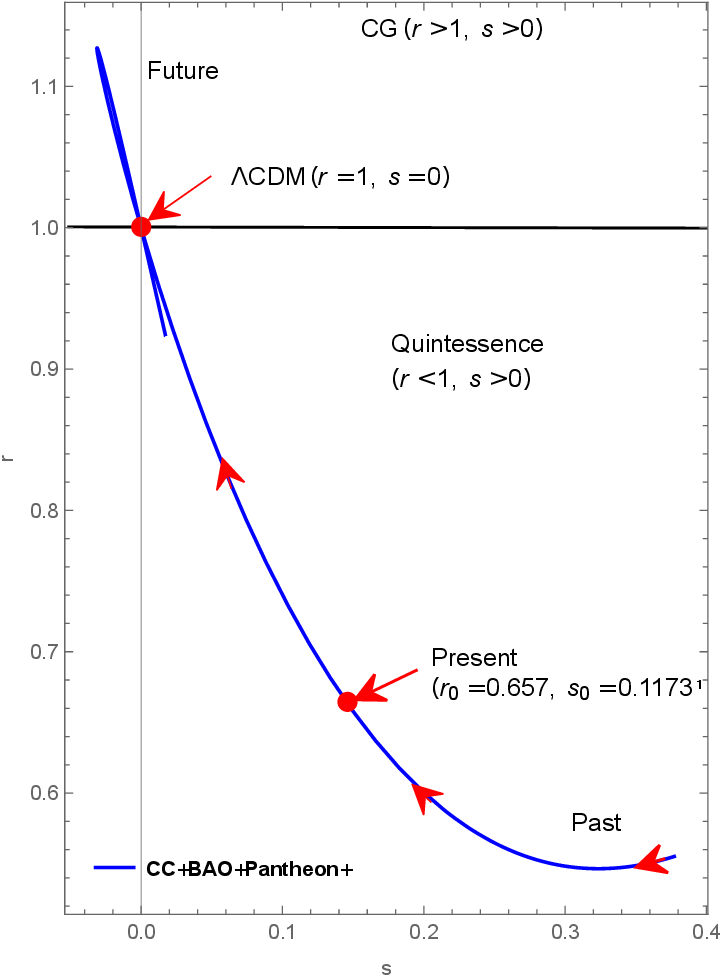}
  \caption{Evolution of the statefinder parameters $r(z)$ and $s(z)$ in the $(r, s)$ plane.}\label{fig:f7}
\end{figure}

From Figure \ref{fig:f7}, we observe a rich dynamical evolution in the $(r, s)$ plane. The trajectory begins at $r=0.3876$, $s=0.555$, which is significantly distant from the $\Lambda$CDM fixed point $(1,0)$. These values correspond to a matter-dominated decelerating phase, where the contribution of the Galileon field is negligible and the cosmic expansion is primarily governed by effective matter-like dynamics. As the Universe evolves, the trajectory moves upward in the $(r, s)$ plane, crossing the $\Lambda$CDM point $(1,0)$, which corresponds to a perfect cosmological constant behavior. This crossing marks a transition from matter-like to dark energy-dominated expansion, which signifies the onset of late-time acceleration. The trajectory continues beyond the $ \Lambda$CDM point, reaching a maximum deviation at $r=-0.0037$, $s= 1.128$. These values signify a transient phantom-like phase or a non-standard acceleration regime, which may be driven by the complex dynamics associated with the Galileon field. After reaching this peak, the curve turns back and gradually converges toward the $\Lambda$CDM fixed point, indicating a stabilization of the expansion rate. This final behavior suggests that the Universe asymptotically approaches a de Sitter-like accelerated phase. At present, the statefinder values are found to be $r_0 = 0.657$ and $ s_0 = 0.1173$, which places the model in the quintessence-like region of the $(r, s)$ diagram.
\section{Om diagnostic}\label{sec6}
\hspace{0.5cm} The $Om(z)$ parameter is a model-independent diagnostic used to differentiate among various dark energy models and to test for possible deviations from the $\Lambda$CDM scenario. Unlike geometrical diagnostics that involve higher-order derivatives of the scale factor—such as the statefinder—the Om diagnostic relies solely on the first derivative of the Hubble parameter, making it especially well-suited for comparisons with observational data. The Om parameter is defined as:
\begin{equation}\label{51}
Om(z)=\frac{\frac{H^{2}(z)}{H_{0}^{2}}-1}{(1+z)^{3}-1},
\end{equation}
In the standard $\Lambda$CDM model, $Om(z)$ remains constant and equals the present matter density parameter $\Omega_{m0}$. Any deviation from this constancy suggests the presence of evolving dark energy. Specifically, $Om(z)> \Omega_{m0}$ indicates phantom behavior $(\omega<-1)$, while $Om(z)<\Omega_{m0}$ suggests a quintessence-like behavior $(\omega>-1)$. Using our Hubble parameter expression $H(z)$, the Om diagnostic becomes:
\begin{equation}\label{52}
Om(z)=\frac{\Omega_{m0}(1+z)^3+(1 - \Omega_{m0})(1+z)^{3(1+\omega_0+\omega_a)} \exp\left(-\frac{3 \omega_a z}{1+z} \right)-1}{(1+z)^{3}-1}.
\end{equation}
\begin{figure}[hbt!]
  \centering
  \includegraphics[scale=0.44]{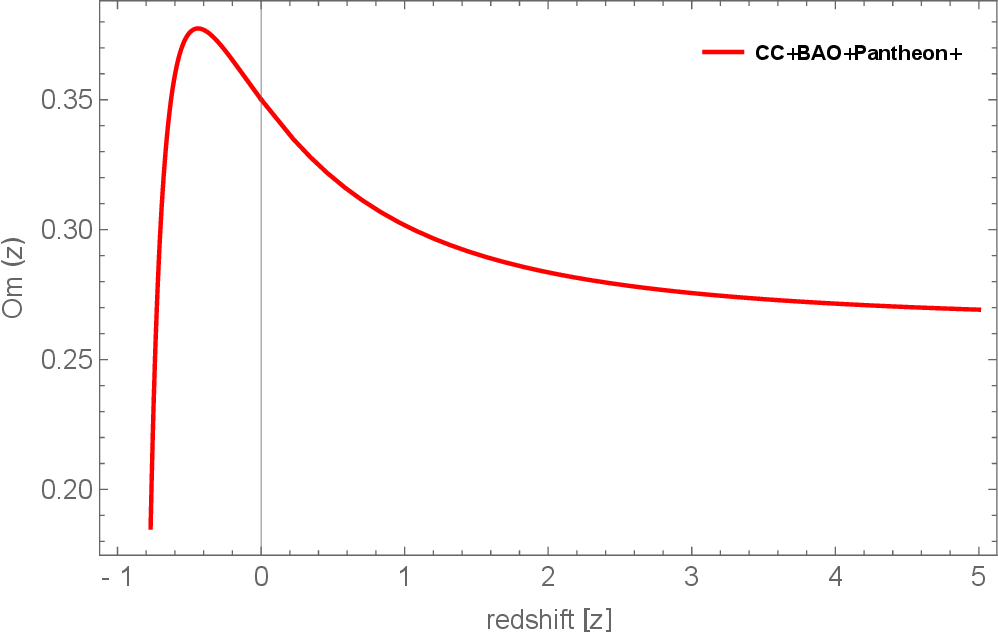}
  \caption{Evolution of the Om diagnostic as a function of redshift.}\label{fig:f8}
\end{figure}

Figure \ref{fig:f8} illustrates the redshift-dependent behavior of the $Om(z)$ parameter. From Figure \ref{fig:f8}, we observe the following behavior: At high redshift, the curve starts with a value of $Om(z)\approx0.27$, which is close to the expected matter density parameter, which indicates a Universe dominated by effective matter-like components. As redshift decreases, the curve increases, reaching a maximum value at redshift $z=0.377$, where $Om(z)=-0.45$. This peak deviation signals a strong departure from standard $\Lambda$CDM behavior and corresponds to a phantom-like phase in the model. After this point, the curve begins to decrease and asymptotically converges toward $Om(z) = -1$ at late times, mimicking a de Sitter-like expansion or cosmological constant behavior. This evolution reflects the dynamical nature of dark energy in our model—transitioning through various effective phases and ultimately settling into behavior consistent with $\Lambda$CDM.
\section{Conclusion}\label{sec7}
\hspace{0.5cm} In this work, we have explored the cosmological implications of a Galileon gravity framework by proposing a parametric form of the Hubble parameter, $H(z)$, motivated by the CPL parameterization of the dark energy equation of state, $\omega_{DE}(z)=\omega_{0}+\omega_{a}\frac{z}{1+z}$. This approach enables a dynamic description of the expansion history within the context of the inherently nonlinear Galileon-modified Friedmann equations. 

Using a combined dataset comprising $4$6 Hubble measurements, DESI DR2 BAO and $1701$ Pantheon+ SNe Ia, we constrained the model parameters through MCMC analysis. The best-fit values obtained are: 
$H_{0} = 67.7043^{+1.4354}_{-1.4102}\,\text{km/s/Mpc}$, $\Omega_{m0} = 0.2668^{+0.0212}_{-0.0217}$, $\omega_0 = -0.8827^{+0.1076}_{-0.0967}$ and $\omega_a = 0.0011^{+0.0660}_{-0.0622}$. 
The model comparison through information criteria yields $\Delta AIC=1.46$ and $\Delta BIC=11.5$, which suggests that our Galileon cosmology is statistically competitive with the $\Lambda$CDM model under AIC, while BIC slightly favors $\Lambda$CDM due to its stronger penalty for model complexity.

We then investigated the evolution of several key cosmological parameters. The deceleration parameter $q(z)$ confirms a transition from deceleration to acceleration at $z_{tr} = 0.7873$, with a present value of $q_0=-0.598$, consistent with an accelerating Universe. Analysis of the energy density and pressure showed that $\rho_{de}(z) > 0$ and $p_{de}(z) < 0$ at all times, which supports the physical viability and dark energy nature of the model.

The equation of state parameter $\omega(z)$ transitions from a matter-like behavior at early times to a quintessence-like regime, ultimately approaching the cosmological constant limit $\omega = -1$. The current value $\omega_0=-0.2915$ suggests that the dark energy component exhibits mild dynamical behavior. Energy condition analysis revealed that the NEC and DEC are satisfied throughout the cosmic history, while the SEC is violated, in line with expectations for models that drive late-time cosmic acceleration.

The statefinder diagnostic demonstrated that the model evolves through distinct dynamical phases— beginning in a matter-dominated regime, which crosses the $\Lambda$CDM fixed point, experiencing a transient phantom-like episode and eventually stabilizing toward a de Sitter-like future. Present-day statefinder values $r_0 = 0.657$ and $s_0 = 0.1173$ place the model within the quintessence region, close to $\Lambda$CDM but allowing for subtle deviations.

Lastly, the Om diagnostic further supports the dynamical nature of dark energy in this model. Starting at $Om(z) \approx 0.27$ at high redshift, the parameter peaks at $Om(z) = -0.45$ around $z = 0.377$, before declining and converging to $Om(z) = -1$ at late times, again reflecting a transition from effective matter-like behavior to a cosmological constant-dominated phase.

Overall, our findings show that the Galileon model with the chosen Hubble parameterization not only fits current observational data well but also provides a rich and consistent description of the Universe’s expansion history. While it aligns with $\Lambda$CDM at late times, it offers a broader dynamical structure, potentially capturing small-scale or early-time deviations that might be revealed with future high-precision cosmological observations.

\end{document}